\documentclass[12pt,preprint]{aastex}
\newcommand{\newdagger}{\dagger}

\shorttitle{EVOLVED GALAXIES AT $z> 1.5$}
\shortauthors{MCCARTHY ET AL.}
\usepackage{multirow}
\begin{document}

\title {Evolved Galaxies at $z > 1.5$ from the Gemini Deep Deep Survey: The
Formation Epoch of Massive Stellar Systems}

\author{ Patrick J. McCarthy$^1$, Damien Le Borgne$^2$, David Crampton$^3$, Hsiao-Wen
Chen$^{5,10}$, \\ Roberto G. Abraham$^2$, Karl Glazebrook$^4$, 
Sandra Savaglio$^{4,9}$, Raymond G.  Carlberg$^2$,\\ Ronald O. 
Marzke$^8$, Kathy Roth$^6$, Inger J{\o}rgensen$^6$, Isobel Hook$^7$, \\
Richard Murowinski$^3$ \& Stephanie Juneau$^3$}

\altaffiltext{1}{Carnegie Observatories, 813 Santa Barbara St,
        Pasadena, CA 91101, pmc2@ociw.edu}

\altaffiltext{2}{Department of Astronomy \& Astrophysics, University of Toronto,
        Toronto ON, M5S~3H8 Canada, leborgne@astro.utoronto.ca, abraham@astro.utoronto.ca, carlberg@astro.utoronto.ca}

\altaffiltext{3}{NRC Herzberg Institute for Astrophysics, 5071 W. Saanich Rd.,
Victoria, BC, Canada, david.crampton@nrc-cnrc.gc.ca, 
murowinski@nrc-cnrc.gc.ca, stephanie.juneau@nrc-cnrc.gc.ca}

\altaffiltext{4}{Department of Physics \& Astronomy, Johns Hopkins
University, Baltimore, MD 21218, kgb@pha.jhu.edu, savaglio@pha.jhu.edu}

\altaffiltext{5}{Center for Space Sciences, Massachusetts Institute of
Technology, 70 Vassar St., Bld. 37, Cambridge, MA 02139, hchen@space.mit.edu}

\altaffiltext{6}{Gemini Northern Operations Center, 670 N. A'ohoku Place,
Hilo, HI 97620, jorgensen@gemini.edu, kroth@gemini.edu}

\altaffiltext{7}{UK Gemini Operations Center, Kebble Road, Oxford
University, Oxford, OX 3RH, UK,}

\altaffiltext{8}{Department of Astronomy and Physics, San Francisco
State University, San Francisco, CA, 94132, marzke@stars.sfsu.edu }

\altaffiltext{9}{On leave of absence from INAF Osservatorio Astronomico di Roma, Italy}

\altaffiltext{10}{Hubble Fellow}

\begin{abstract}

    We present spectroscopic evidence from the Gemini Deep Deep Survey (GDDS)
for a significant population of color-selected red
galaxies at $1.3 < z <  2.2$ whose integrated light is dominated by evolved stars.
Unlike radio-selected objects, the $z > 1.5$ old 
galaxies have a sky density $> 0.1$ arcmin$^{-2}$.  Conservative 
age estimates for 20 galaxies with $z > 1.3; \langle z \rangle =
1.49$, give a median age of 1.2~Gyr and $\langle z_f \rangle = 2.4$.  
One quarter of the galaxies have inferred $z_f > 4$.  
Models restricted to [Fe/H] $\leq 0$ give median ages and $z_f$ of 2.3~Gyr and 
3.3, respectively.  These galaxies are among the most massive and contribute $\sim 50$\% of the 
stellar mass density at $1 < z < 2$.  
The derived ages and most probable star formation histories suggest a 
high star-formation-rate ($\sim 300-500~\mathrm{M}_{\odot}\mathrm{yr}^{-1}$) 
phase in the progenitor population. We argue that most of the red galaxies are 
not descendants of the typical $z \sim 3$ Lyman break galaxies.  Galaxies associated with
luminous sub-mm sources have the requisite star formation rates
to be the progenitor population.  Our results point toward early and rapid 
formation for a significant fraction of present day massive galaxies.

\end{abstract}


\section{Introduction}

Recent surveys have placed significant constraints on galaxy formation models. 
The evolving stellar mass density (e.g. Glazebrook et al. 2004, hereafter Paper III; 
Dickinson et al. 2003; Rudnick et al. 2004; Fontana et al.
2004) now seems in reasonable accord with measurements of integrated star formation rates 
(e.g. Steidel et al. 1999). Interpretation of these volume-averaged 
quantities remains difficult, in part because the earliest phases of galaxy formation are 
poorly understood. 
Age determinations for galaxies at intermediate and high 
redshift provide an accurate clock for galaxy formation at early times. 
Application of this technique has been hampered by the small 
samples of suitable galaxies at appropriate redshifts, and the 
difficulty of obtaining reliable age estimates from low signal-to-noise spectra.
Age determinations for the prototypical evolved red galaxy, 53W091,
range from $> 3.5$~Gyr (Dunlop et al. 1999; Nolan et al. 2001) 
to as young as $1-1.5$~Gyr (Bruzual \& Magris 1997; Yi et al. 2000).  
Visible-light surveys of galaxies at $z > 1$ are biased against inclusion of 
red galaxies and hence are limited in their ability to shed light on the formation
of the oldest and most massive galaxies. Near-IR surveys of galaxies with red 
optical-to-IR colors provide the requisite samples of massive galaxies at 
redshifts that constrain the formation epoch. 

Most spectroscopic studies of the red $R-K$ or $I-K$ population to date
(e.g. Cimatti et al. 2002; Yan et al. 2004), 
have not revealed signatures of evolved 
populations at z $ > 1.3$ due to observational limitations. At higher
redshifts the strong signatures of older stellar populations (e.g.,
4000~\AA{} break and CaII H\&K lines) are beyond the reach of most CCD-based
spectrographs.  At z $> 2$,  strong UV resonance lines 
are redshifted into the atmospheric passband, allowing identification 
of UV-bright galaxies.  In the intermediate regime, $1.3 < z < 2$, evolved 
galaxies must be probed with weak photospheric features of MgII, MgI and FeII.

The red galaxy population \footnote{A review
of the red near-IR selected population can be found in McCarthy (2004).}
(e.g. $I-K > 4$) is a mix of evolved and reddened star forming systems. 
Estimates of the fraction of star forming galaxies in this population
range from 20 - 60\% (e.g. Cimatti et al. 2002; Smail et al. 2002; Yan et al. 2004).
Although the $z > 2$ red $J-K$ selected galaxy population 
(e.g. Franx et al. 2003) likely contains evolved objects, 
those with confirmed redshifts all have strong star formation signatures in their rest-UV 
spectra (van Dokkum et al. 2003), possibly due to observational
limitations. Similarly, the massive K-bright galaxies at $z \sim 2$
from the GOODS and K20 surveys (Daddi et al. 2004) have high inferred
star formation rates. The Gemini Deep Deep Survey (GDDS: Abraham et al. 2004, hereafter Paper I), 
was the first to identify evolved galaxies at $z > 1.5$ in significant numbers (also see
Cimatti et al. 2004).

The GDDS is a spectroscopic study of galaxies in the $1 < z < 2$ range 
within four separate areas of the Las Campanas IR Survey (McCarthy et
al. 2001). Very long exposures, using the `Nod \& Shuffle' sky cancellation technique 
yielded high quality spectra from
which redshifts could be derived for 272 galaxies with $I < 24.5$ (Vega).
Catalogs, field locations, mask design,  and sample
selection are given in Paper I. Twenty GDDS galaxies with
$z>1.3$ display clear signatures of old stellar populations. In
this paper we present spectra of these galaxies along with 
preliminary age determinations and 
consider the implications of these results for our understanding of 
the formation of massive galaxies.

\section {Quiescent Red Galaxies at $z > 1.3$}

The sample of 20 GDDS galaxies with $z >1.3$ with spectra
characteristic of old populations is listed in
Table~1. The object designation is given in the first column,
followed by the redshift. Confidence classes and spectral classifications are
given in Paper I. High confidence
redshifts were determined for 71\% (75/105) of the $I-K>3.5$ galaxies and 
67\% (35/52) of the $I-K>4$ subsample. 
These two samples span the redshift range from 0.8-2.1. 
Among objects with spectroscopic redshifts, 51\% of the $I-K>3.5$ sample have
unambiguous old stellar components. The redder galaxies ($I-K>4$) show an even
greater preponderance for pure-old spectra with 72\% having significant old
populations.  About half of the objects with $I-K>3.5$ and  $z > 1.3$ 
have old spectra, while 80\% of the objects with $I-K>4$ and $z > 1.3$ have old spectral
classifications.
{\it Thus 50-80\% of the $z>1.3$ red GDDS sample show spectra with
contributions from old stellar populations}. The contribution
of old stars in our $z > 1.3$ sample is similar to that seen
in the $z \sim 1- 1.3$ samples (e.g., Cimatti et al. 2002; Yan et al. 2004).


\begin{table}[htbf]
\caption{Best Fitting Ages for Red Galaxies.\smallskip}
\begin{tabular}{cccccccccrr}
\hline
\hline
 & & & & & \multicolumn{5}{c}{Model Parameters} &\\
\cline{6-10}
Object&
$z$   &
K     &
I-K   &
Mass\tablenotemark{\newdagger}  &
Age(Gyr) &
$\tau$\tablenotemark{\newdagger\newdagger} &
A$_{\rm v}$ &
Z/Z$_{\sun}$ &
$\chi^2$ &
z$_f$ \\
\hline
12-6131 & 1.308 &  19.2  & 4.2 & 10.9 & 0.8$_{-0.3}^{+0.6}$  & 0.0 & 0.8 & 1.00 & 1.4 & 1.6   \\
02-1255 & 1.340 &  18.3  & 4.7 & 11.2 & 1.0$_{-0.3}^{+1.7}$  & 0.0 & 0.8 & 1.75 & 1.9 & 1.8   \\
02-1842 & 1.342 &  18.7  & 4.3 & 10.9 & 0.9$_{-0.2}^{+0.7}$  & 0.0 & 0.6 & 1.75 & 1.9 & 1.7   \\
12-5836 & 1.348 &  18.9  & 3.6 & 10.7 & 0.5$_{-0.1}^{+0.1}$  & 0.0 & 0.6 & 1.30 & 3.6 & 1.5   \\
15-7972 & 1.361 &  19.1  & 4.0 & 10.8 & 2.0$_{-1.4}^{+2.5}$  & 0.0 & 0.2 & 0.30 & 1.6 & 2.2   \\
22-2587 & 1.395 &  19.3  & 4.1 & 10.7 & 1.5$_{-0.8}^{+2.9}$  & 0.0 & 0.0 & 1.00 & 1.6 & 2.5   \\
22-0948 & 1.396 &  18.9  & 4.3 & 10.8 & 4.0$_{-3.5}^{+0.4}$  & 0.5 & 0.0 & 0.75 & 1.6 & $>5$  \\
12-8025 & 1.397 &  18.9  & 4.2 & 11.1 & 0.8$_{-0.1}^{+0.6}$  & 0.0 & 0.0 & 1.75 & 2.2 & 1.8   \\
22-0107 & 1.450 &  18.3  & 4.9 & 11.3 & 3.2$_{-2.8}^{+0.7}$  & 0.5 & 0.4 & 0.75 & 1.3 & $>5$  \\
22-1983 & 1.488 &  19.1  & 4.6 & 11.1 & 1.1$_{-0.5}^{+3.1}$  & 0.0 & 0.0 & 1.30 & 1.3 & 2.1   \\
22-0189 & 1.490 &  18.1  & 4.8 & 11.5 & 3.0$_{-0.2}^{+0.7}$  & 0.5 & 0.4 & 0.75 & 1.9 & 4.8  \\
22-0674 & 1.493 &  18.8  & 4.4 & 11.1 & 3.4$_{-1.7}^{+0.3}$  & 0.5 & 0.0 & 0.75 & 1.6 & $>5$  \\
12-5869 & 1.510 &  18.6  & 2.6 & 11.5 & 1.2$_{-0.2}^{+0.6}$  & 0.0 & 0.8 & 0.25 & 2.0 & 2.2   \\
12-6072 & 1.576 &  19.8  & 4.3 & 10.8 & 1.6$_{-1.3}^{+2.1}$  & 0.2 & 0.2 & 0.75 & 2.0 & 2.7   \\
12-5592 & 1.623 &  19.4  & 3.9 & 11.1 & 1.1$_{-0.4}^{+0.3}$  & 0.1 & 0.0 & 0.75 & 3.0 & 2.3   \\
12-8895 & 1.646 &  18.5  & 4.7 & 11.5 & 2.6$_{-0.3}^{+0.3}$  & 0.5 & 0.4 & 0.75 & 2.2 & 4.7  \\
15-4367 & 1.725 &  19.5  & 4.1 & 10.7 & 2.1$_{-0.9}^{+0.4}$  & 0.0 & 0.0 & 0.20 & 2.4 & 3.8  \\
15-7543 & 1.801 &  19.0  & 4.6 & 11.0 & 0.9$_{-0.2}^{+0.5}$  & 0.0 & 0.0 & 1.75 & 1.9 & 2.4   \\
15-5005 & 1.845 &  19.6  & 4.0 & 10.8 & 0.5$_{-0.1}^{+0.7}$  & 0.1 & 0.2 & 0.74 & 2.4 & 2.2  \\
12-7672 & 2.147 &  19.1  & 4.4 & 11.1 & 1.2$_{-0.4}^{+0.1}$  & 0.1 & 0.0 & 0.74 & 2.0 & 3.4   \\
\hline
\end{tabular}
\tablenotetext{\newdagger}{$\log_{10} \mathrm{M}_\mathrm{*}/\mathrm{M}_\odot$ from Glazebrook et al.
 2004 (Paper III)}
\tablenotetext{\newdagger\newdagger}{$e$-folding time of star formation rate in Gyr.}
\end{table}


\begin{figure}
\includegraphics[width=18cm]{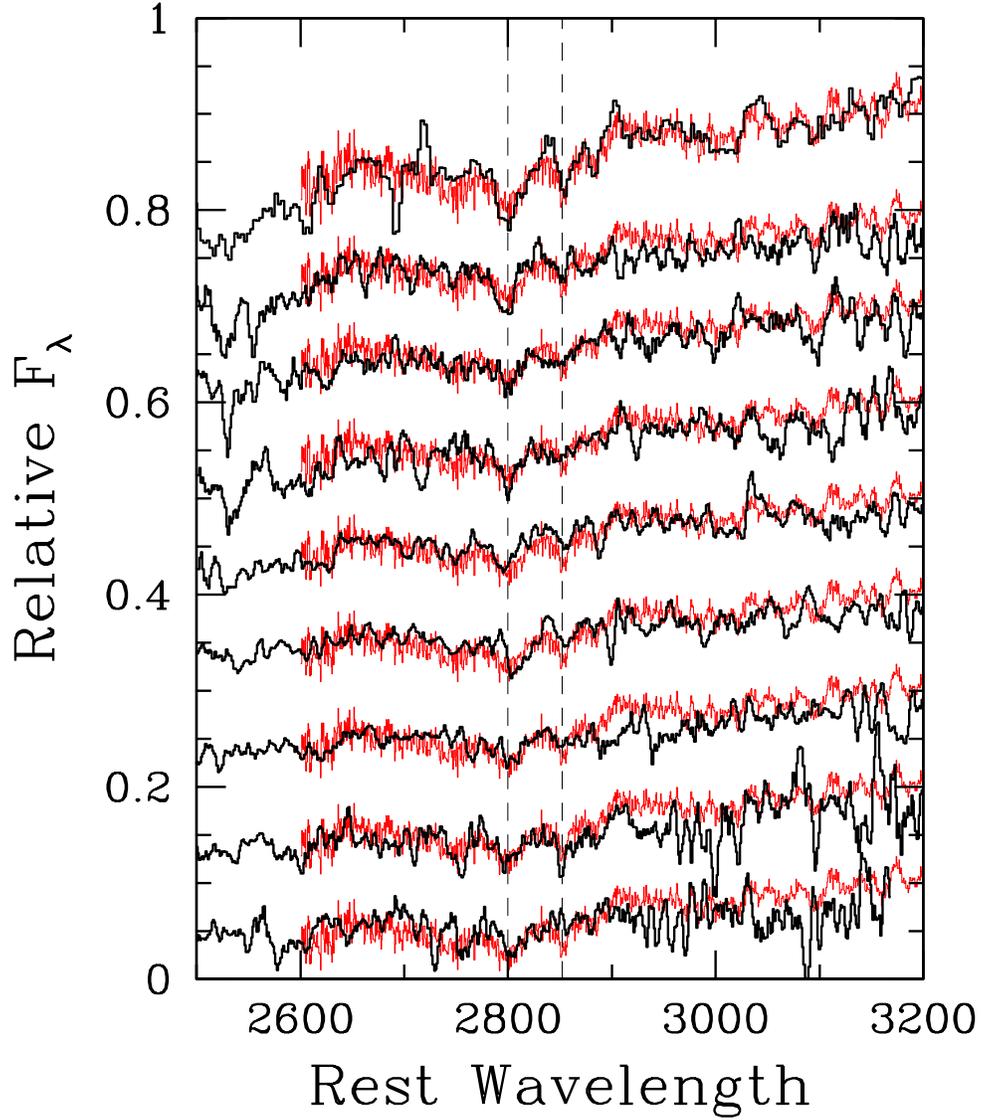}
\caption{Spectra of evolved GDDS galaxies with $z > 1.3$. From
top to bottom the objects shown are: GDDS-02-1255, 22-0189, 22-0674,
12-5869, 12-6072, 12-8895, 15-4367, 15-7543,15-5005.  This includes
all of the galaxies in Table~1 with $1.49 < z < 2.0$, plus
GDDS-02-1255 ($z = 1.34$).  The SDSS LRG composite has been overlaid
on each spectrum and an offset has been applied to each, in steps of
$10^{-18}$~erg s$^{-1}$ cm$^{-2}$ \AA$^{-1}$.  The locations of the
stellar MgII2800 and MgI2852 lines are indicated by the dashed lines.}
\end{figure}

\begin{figure}
\includegraphics[width=15cm]{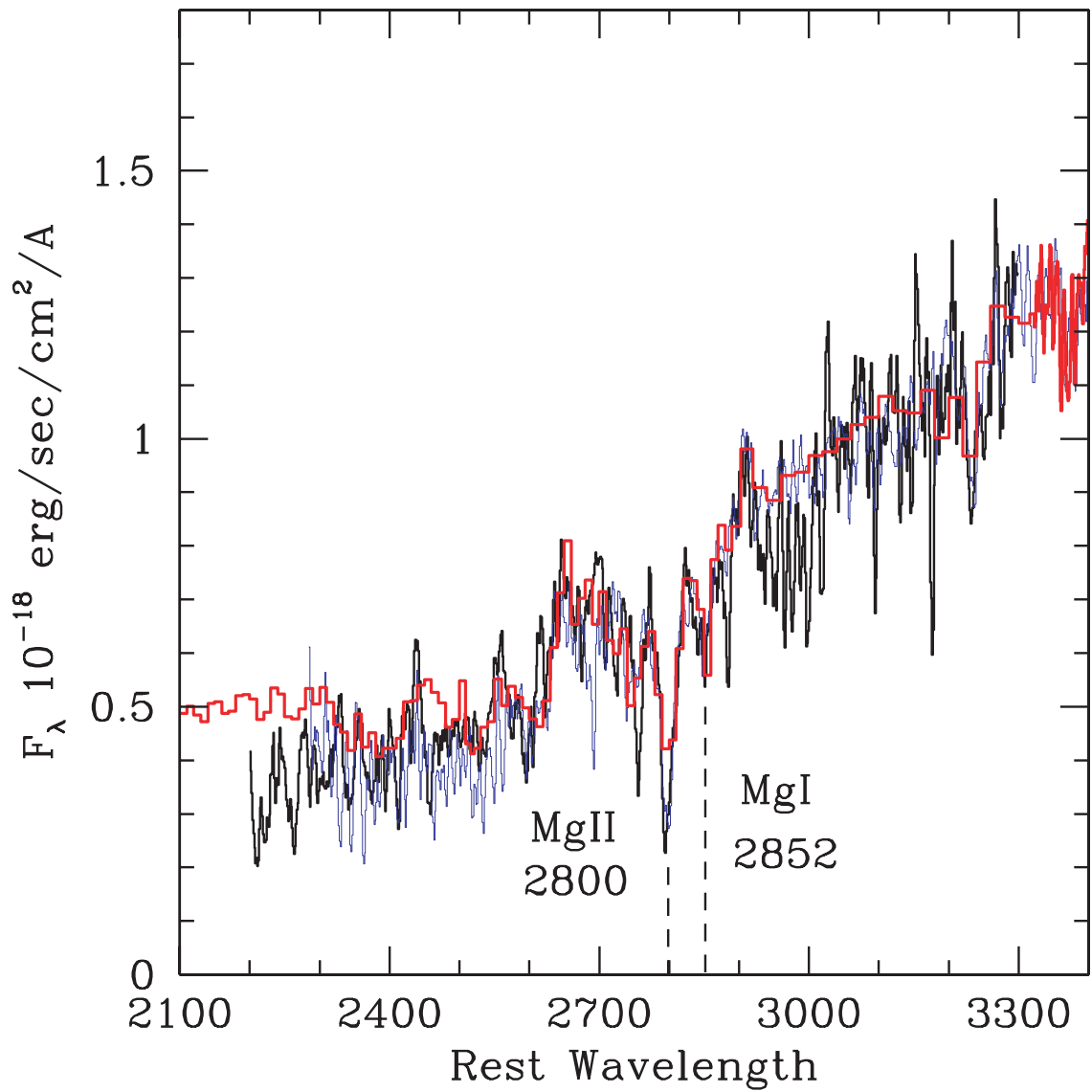}
\caption{Composite spectra of evolved galaxies with $1.3 < z <
1.4$ (in blue, from galaxies  GDDS-12-6131, 02-1255, 12-5836, 15-7972, 22-2587, and
12-8025), and $1.6 < z < 1.9$ (in black, from the five objects
listed in Table~1 in this redshift range). Both composite
spectra show strong MgII2800, MgI2852 absorption and broad spectral
features due primarily to FeII absorption. Overlaid in red is a
single-burst Bruzual \& Charlot (2003) spectral synthesis model with an age of 2~Gyr,
solar abundances, and a Salpeter (1955) IMF cutoff at $120M_{\odot}$.
}
\end{figure}


\subsection {Individual and Composite Spectra}

   In Figure~1 we show spectra of 9 $z >
1.3$ objects with high-confidence redshifts and spectral classes indicative of 
old populations from the four
GDDS fields.  Over-plotted on these spectra is the SDSS/LRG template
(Eisenstein et al. 2003). The evolved GDDS
galaxies form a fairly homogeneous set and are reasonably well
matched to the SDSS/LRG template. The strongest features in the
spectra are the MgII2800 doublet and MgI2852. Numerous weak FeII
features blend together to produce a modulated shape to the continuum.
The overall spectral slopes and shapes of the objects with $z = 1.6 -
1.8$ are not very different from those at $z = 1.3$, although they
have somewhat flatter spectral slopes at $\lambda < 2800$\AA.

   In Figure~2 we show composite spectra of six galaxies from Table~1
with $1.3 < z < 1.4$, and another composite of the five at $1.6 < z <
1.9$.  The equivalent widths of the MgII and MgI lines and
overall continuum shapes are quite similar. A spectral synthesis model derived using 
Bruzual \& Charlot (2003) is shown over-plotted in red in
Figure~2. This reddening-free single-burst simple stellar population model, 
with an age of 2~Gyr, a Salpeter (1955) IMF and solar abundances, is
the youngest solar-abundance model that fits the data well. Older models
(e.g. ages of $3-4$~Gyr) fit the far-UV end of the spectrum better than the
2~Gyr model does, but not at a level that allows us to rule out the
2~Gyr model.  Younger models (e.g., 1~Gyr) are
bluer than the observed composite spectra.  The characteristic ages
derived from the composite spectra are consistent with
the results from fits to the spectra and broad-band energy
distributions of individual objects (Sec.~3.1).

Figures~1 and 2 convey the key result presented in this paper: there is a
significant population of color-selected luminous field galaxies at 
$1.3 < z < 2$  with spectra dominated by old stars. 
Unlike radio-selected evolved galaxies at $z \sim 1.5$ previously reported
(e.g. Dunlop et al. 1996), our objects have a high
surface density on the sky ($> 0.1$ sq. arcmin$^{-2}$).

\section {Model Ages and Formation Redshifts}

The existence of massive galaxies at redshifts up to $z \sim 2$
with evolved spectra argues for an early formation episode and strains
semi-analytic CDM models (e.g. Cole et al. 2000; Baugh et al. 2003). The
most successful models (Somerville et al. 2004) can produce sufficiently
massive galaxies at early epochs, but still have
difficulties in producing red and old galaxies at $z>1$.  The critical
empirical issue is understanding the age of the stellar populations,
and hence the formation redshifts, in the old red galaxies at the highest redshifts possible.

We made a preliminary attempt to derive very conservative (i.e. minimum) 
ages for the galaxies listed in Table~1 by making use of both the information in the spectra and
$B - K_s$ photometry.  For each galaxy, we systematically compared the observed SED with a
set of synthetic spectra computed with P\'EGASE.2 (Fioc \&
Rocca-Volmerange 1997) and constructed a multi-dimensional
$\chi^2$ surface spanning a wide range of star-formation histories,
ages, extinction (A$_{\rm v}$) and metallicities.

The observed SED of each galaxy comprises a flux-calibrated GDDS spectrum and a
broad-band spectral energy distribution combined
with weights assigned in proportion to the band-width of the
observations.  In most cases the spectra and broad-band photometry
carry nearly equal weight.  All the models use a Salpeter (1955) IMF with an upper-mass
cutoff of 120M$_{\odot}$. The effects of reddening were modeled by using
the Calzetti (1997) extinction law with A$_{\rm v}= 0-1$mag.
Instantaneous-burst models with 12~metallicities ranging from 20\% to
175\% of solar were considered, together with exponentially
declining star formation histories with 10~e-folding times ranging
between 0.1~Gyr and 3~Gyr.

From the $\chi^2$ surface, a best-fit age, star formation history,
metallicity and extinction were derived.  The range of
acceptable ages for a given galaxy was limited by the age of the
universe at its observed redshift. Statistical uncertainties on the age were computed 
from the set of models satisfying $\Delta \chi^2 < 3 \sigma$.
An extensive description of the underlying models and our approach to
fitting data to these will be given in Le Borgne et al. (in preparation).

\subsection {Results}

We summarize the results of our age determination analysis using the
P\'EGASE.2 models in Table ~1. 
In column 6 of Table~1 we list the best-fit ages and the range of acceptable
ages.  Bruzual \& Charlot (2003)
instantaneous-burst models yield best fits with age differences smaller than 0.2~Gyr.
Columns 7 through 11 give the e-folding time, A$_{\rm V}$, abundance,
$\chi^2$ value and formation redshift for the best-fitting models,
respectively. The reduced $\chi^2$ values exceed unity, in part, because of imperfections
in the model spectral libraries.  The formation redshifts are determined from the
redshift of observation, the best-fit age and the age of the Universe
in a $\Omega_\Lambda = 0.7, \Omega=0.3, H_0 = 70$~km s$^{-1}$Mpc$^{-1}$ cosmology. 
In nearly all cases the best fits
were achieved with either an instantaneous burst or a short e-folding time exponential
burst ($<0.5$~Gyr), with a preponderance favoring the instantaneous-burst models.

The single-burst and exponential model fits generally
favor metallicities higher than 50\% solar. Super solar metallicities reduced the
$\chi^2$ values in $\sim 30$\% of the cases. We truncated our
metallicity search space at [Fe/H]$=0.25$. While large values are seen in the cores
of local ellipticals (e.g. Thomas et al. 2003), the large apertures of our spectroscopic
and photometric measurements ($10-20$kpc diameter) and the abundance gradients seen in
ellipticals produce luminosity weighted abundances within our apertures that are closer
to solar (see J{\o}rgensen et al. 1997; Arimoto et al. 1997).
For the four objects that favor 
models with [Fe/H]$>0.15$, fits with the metallicity capped at the solar value 
yield best-fit ages that are 0.5-1.0~Gyr older than those listed in Table~1.

We have measured the mass-density in
high-mass objects and determined the contribution from the $I-K >3.5$ objects
using the procedures from Paper III.
Adopting a threshold of $5 \times 10^{10}\mathrm{M}_\odot$ (for which our sample is
mass-complete over the interval $1.3<z<2$) we find that the red objects in this sample
contribute 49\% of the overall stellar mass density.

The median derived age and formation redshifts are
$1.2$~Gyr and $2.4$, respectively for our conservative analysis.
Nearly 1/3 of the objects (5/20) have inferred $z_f > 4$.
Imposing a minimum collapse time of
$3 \times 10^8$~yr moves the median formation redshift from $2.4$ to $\sim 3.0$. Limiting
the models to [Fe/H] $\leq 0$ shifts the inferred median age and $z_f$ to 2.3~Gyr and 3.2,
respectively; imposing A$_{\rm v}$ = 0 in addition yields a median $z_f$ of 4.0.

\section{Discussion}

The spectra of the red GDDS galaxies reveal unambiguous evidence for
old and metal-rich galaxies over the full range from $1 < z < 2$.  
Our preliminary, conservative analysis implies early and rapid formation for a substantial
fraction of these.  The median $z$, age, $z_f$, and mass for the sample in Table~1
are 1.49, 1.5~Gyr, 2.5, and $1 \times 10^{11}$~M$_{\odot}$, respectively.  Our
models strongly favor instantaneous bursts in roughly 50\%
of the objects and short e-folding times (e.g. 0.5~Gyr) for the remaining galaxies.
More plausible models, those with star formation extended over
one or more dynamical times, produce
best-fitting ages that are typically $1$~Gyr $larger$ than those in Table~1, implying
$<z_f> \sim 4$ for a substantial fraction of the galaxies.

Taking the $z_f$ in Table~1 as indicative of the onset of star formation, the ``median'' red
galaxy requires a constant star formation rate of 50~M$_{\odot}$ yr$^{-1}$ over
the full 1.5~Gyr period from $z = 2.4$ to $z = 1.5$ to produce the required stellar
mass.  These rates are higher than that of a typical Lyman Break Galaxy (LBG)
at $z \sim 3$ (e.g. Shapley et al.  2003). Constant star formation rate models, however,
reproduce neither the colors nor the spectra of the red galaxies. 
Star formation with an e-folding time of $3 \times 10^8$~yr implies peak
SFRs $\sim 300-500$~M$_{\odot}$yr$^{-1}$ at $z \sim 2-4$ in the most
massive galaxies. These high star formation rates, coupled with the strong clustering
of the red galaxy population (e.g. McCarthy et al. 2001) and the
differences in stellar mass, suggest that these objects
are probably not closely connected with the $z = 3-4$ Lyman-break galaxy population.

Assembly of the massive GDDS galaxies from many sub-units still requires early, 
and short-lived, star formation, although the units could be smaller than the 
typical $8 \times 10^{10}$M$_{\odot}$ of the GDDS galaxies.  
While star formation in smaller sub-galactic units will proceed with shorter dynamical 
times, the best-fitting instantaneous burst (or even $\tau = 0.1$~Gyr) models imply impressive 
synchronization in the truncation of star formation among the precursors.
The near-solar metallicities required to fit the 
strong UV stellar lines and the favored single-burst models naively argue for
a single massive star formation episode (per galaxy) at $2 < z < 5$ as
the simplest formation scenario. 

It appears that there
is a continuous coeval population of massive red galaxies that are traced,
in order of decreasing redshift, by: $z > 2$ red $J-K$ galaxies, red GDDS
galaxies at $z \sim 1.3-2$ and the classical ``EROs'' at $0.9 < z < 1.3$ observed in the
K20, LCIR and other surveys. These all point to an early formation epoch for
the progenitors of present day massive galaxies. Our analysis implies high peak star
formation rates at $z \geq 2$.  At present the only candidates 
for such rapidly forming massive galaxies at high redshift are the
sub-mm luminous sources.  The median redshift ($\langle z\rangle = 2.4$) for the 
bright  SCUBA sources (Chapman et al. 2003) is indistinguishable from our
current conservative $\langle z_f\rangle$ estimate for the red GDDS galaxies. A significant fraction 
of the GDDS galaxies, however, appear to have formation redshifts outside the range of
known SCUBA redshifts.  The $n(z)$ of the SCUBA sources, however, has an inferred 
tail to $z >4$ and these may evolve into the red $z > 1.5$ population.
In summary, the GDDS has revealed a population of
evolved galaxies at $z > 1.5$ and conservative age estimates yield a modest 
$\langle z_f\rangle$ and point to formation of massive galaxies in episodes of intense 
star formation.

\section {Acknowledgments}
This paper is based on observations obtained at the Gemini
Observatory, which is operated by  AURA Inc., under a cooperative 
agreement with the NSF on behalf of the Gemini partnership: the NSF (US),
PPARC (UK), NRC (Canada), CONICYT (Chile), ARC
(Australia), CNPq (Brazil) and CONICET (Argentina). This paper is also
based on observations obtained at the Las Campanas Observatory of the
Carnegie Institution of Washington. KG and SS acknowledge support from the
David and Lucille Packard Foundation, RGA acknowledges support from the NSERC.

\end {document}